\documentclass[twocolumn,showpacs,amsmath,amssymb,showkeys,
floatfix,prl]{revtex4}
\usepackage{amsfonts,amsmath}
\usepackage{graphicx}
\usepackage{dcolumn}
\usepackage{bm}

\begin{document}

\title{Correlations between $\theta_{13}$ and $\theta_{23}$ in a very long
baseline neutrino oscillation experiment}

\author{D. C. Latimer}
\affiliation{Department of Physics, University of Louisville, 
Louisville, Kentucky  40292, USA}

\author{D. J. Ernst}
\affiliation{Department of Physics and Astronomy, Vanderbilt University, 
Nashville, Tennessee 37235, USA}
\date{\today}

\begin{abstract}
A very long baseline experiment, with $L/E$ of $1.6 \times 10^4$ m/MeV, 
has been proposed as the ideal means to 
measure $\theta_{13}$, including its sign, for neutrino energies less than 50 
MeV. Higher energies require consideration of the earth MSW effect. 
Approximating a constant density mantle, we examine the $\nu_e$--$\nu_\mu$ and 
$\nu_\mu$--$\nu_\mu$ oscillation channels at very long baselines with 
particular focus upon the deviation of $\theta_{23}$ from maximal mixing and 
the sign of $\theta_{13}$.  It is demonstrated that measurements of these 
oscillation channels in this region are sensitive to $\theta_{13}$, including its sign, 
and to $\theta_{23}$, including its octant, with a significant correlation between the 
value of $\theta_{13}$ and the difference of $\theta_{23}$ from maximal mixing. 
\end{abstract}

\maketitle
\section{Introduction}

All existing neutrino experiments (except the LSND \cite{lsnd} 
appearance experiment) can be understood within the context of three-flavor 
neutrino oscillations. By relating three mass eigenstates to the flavor 
states via a unitary transformation parameterized in the canonical manner 
\cite{pdg}, one has six free parameters with which to fit the experimental 
data:  two mass-squared differences, three mixing angles, and one Dirac CP 
phase.  Solar neutrino experiments \cite{solar} along with KamLAND 
\cite{kamland} point to the large mixing angle MSW solution with a 
mass-squared difference $\Delta_{21} \sim 8 \times 10^{-5}$ eV$^2$ (with 
$\Delta_{jk}:=m_j^2-m_k^2$) and mixing angle $\theta_{12} \sim 0.58$. 
The results of the Super-Kamiokande atmospheric neutrino experiment 
\cite{superk}, in addition to the K2K experiment \cite{k2k}, indicate another mass scale
with $\Delta_{31} \sim 2 \times 10^{-3} \mathrm{eV}^2$ and a near maximal 
mixing angle $\theta_{23} \sim 0.79$. To date, of the reactor experiments, 
data from CHOOZ provide the most stringent limit on the magnitude of the 
remaining mixing angle $\vert \theta_{13} \vert < 0.22$ at 90\% CL for 
$\Delta_{31} = 2 \times 10^{-3} \mathrm{eV}^2$ 
\cite{chooz}.  No information is known about the level of CP violation in the 
neutrino sector; hence, there are no limits on the phase $\delta$.
Recent analyses can be found in
Refs.~\cite{MSTV,BGP} and recent reviews in 
Refs.~\cite{bargerrev,neutreview,BP,BK}.

A high-priority goal of the neutrino physics community is to more precisely
determine the value of $\theta_{13}$.
Future reactor experiments  
have been proposed to improve upon the CHOOZ results with the
hope of establishing a lower bound on the magnitude of this mixing angle \cite{reactor}. 
Measuring electron neutrino disappearance from a reactor yields a clean measurement
of $\sin^2 \theta_{13}$; however, this measurement cannont contribute
to determining the sign of the angle. 
In order to reduce the systematic error, these experiments 
will have a near and far detector. An increase in sensitivity to $\sin^2\theta_{13}$
can be achieved by an innovative use of the near detector \cite{sergio1}. In addition, for 
atmospheric neutrinos traveling through the earth, the ratio of multi-GeV $e$-like to 
$\mu$-like fluxes is sensitive to $\sin^2\theta_{13}$ as well as the 
mass hierarchy \cite{sergio2}.  Multi-GeV atmospheric neutrinos can also provide 
information on $\sin^2\theta_{13}$ and the hierarchy question through a measurement of 
the difference between $\mu^+$ and $\mu^-$ events \cite{sergio3,DIMVNM}. 
Such a measurement would require an 
iron calorimeter detector such as the MINOS detector \cite{minos}.

We are particularly interested in measurements that are sensitive to terms in the oscillation
probabilities which are linear in  $\sin\theta_{13}$. 
It is possible that experiments designed to focus upon such measurements might 
further reduce the errors on the value of $\theta_{13}$. In addition, other information about the mixing
angle can be garnered from these measurements.
We adopt the choice of bounds on the mixing angles and CP phase as proposed in
Ref.~\cite{angles}.  In contrast to the standard convention, this choice restricts the CP
phase to $0 \le \delta < \pi$ while extending one mixing angle to negative values
$-\pi/2 \le \theta_{13} \le \pi/2$.  The advantage is that the parameter space is a
connected region whenever one takes CP to be conserved.  Given this choice, it is clear that terms
in the oscillation probability which are linear in $\sin \theta_{13}$ can aid in determining the
sign of this mixing angle.  

The exact expressions for the oscillation probabilities necessarily exhibit terms linear
in $\sin \theta_{13}$ \cite{bargerrev}; however, given that oscillations occur at two different mass
scales, most current experiments can be understood to a good approximation within the context of modified
two-neutrino oscillations.  In Ref.~\cite{expansion}, one finds perturbative expansions about the
small quantities $\theta_{13}$ and/or the ratio of the mass-squared differences $\alpha :=
\Delta_{21}/\Delta_{31}\sim 0.03$.  From these expansions, we note that terms linear in $\theta_{13}$ are
supressed by a factor of $\alpha$.  Clearly, this small mixing angle will be of little consequence
whenever this perturbative expansion is valid.  Whenever oscillations due to the small
mass-squared difference interferes with oscillations due to the large mass squared difference, the expansion 
is no longer a good approximation at low
orders \cite{expansion}.  In terms of the baseline $L$ and neutrino energy $E$, these subdominant oscillations need to be
considered whenever $L/E \gtrsim 10^4$ m/MeV.

The sub-GeV sample of atmospheric neutrinos in the Super-K experiment contains such baselines \cite{superk}.  
The implications of these subdominant oscillations
for the analysis of atmospheric neutrinos has been thoroughly investigated in Ref.~\cite{smirnov1}.   
Terms linear in $\theta_{13}$, in part, could explain a possible excess of 
$e$-like events in the Super-K data. Earlier
work examined this excess but in the absence of the linear terms \cite{eexcess}.
Additionally, in Ref.~\cite{CGMM}, a full three neutrino analysis of the atmospheric data is presented. 
In this work, 
the authors assume CP conservation; as they allow only positive $\theta_{13}$, they must consider
two CP phases  $\delta = 0, \pi$.  Relating this to our choice, negative $\theta_{13}$ corresponds
to $\delta= \pi$.  Small differences between the two choices of the CP phase are a direct result of
the terms linear in $\sin\theta_{13}$.  Finally, in the analyses presented in \cite{th13exp,us}, the
authors note a small asymmetry in $\chi^2$ about $\theta_{13}=0$ attributable to such linear terms. 
Though statistically insignificant, two local minima in $\chi^2$ were found, one for positive
$\theta_{13}$ and one
negative.

In Ref.~\cite{th13exp}, it was determined, in vacuo, that the $\nu_e$--$\nu_\mu$ and
$\nu_\mu$--$\nu_\mu$ oscillation channels exhibited signficant linear dependence upon $\theta_{13}$
whenever one had a baseline-energy ratio of $L/E =2(2n+1)\pi/\Delta_{12}$.
With $n=0$, the first $\Delta_{21}$ oscillation trough, a twenty-five percent effect was found on the
$\nu_\mu$--$\nu_\mu$ channel for the presently allowed range of $\theta_{13}$.  This ideal baseline
is $L/E = 1.6 \times 10^4$ m/MeV.

A purpose of this work is to extend the results of Ref.~\cite{th13exp} to include 
the MSW effect \cite{msw} so that one may employ higher energy neutrinos to probe the region of
interest cited above. We show herein that even with higher energy neutrinos an experiment which is
sensitive to linear terms in $\theta_{13}$
is still beyond 
present technology but perhaps more feasible. 
We also examine the correlation 
between the value of $\theta_{13}$ and the deviation of $\theta_{23}$ from maximal 
mixing.  We take as known all parameters but these two mixing angles; granted, a full consideration
of all parameters can have significant impact upon such correlations \cite{freund}.
We first consider vacuum oscillations which showcase the essential feautres. 
In the subsequent section, we permit matter interactions via the MSW effect.  An approximate
analytical expression is derived to demonstrate the relation between the two mixing angles.
The general features found in the vacuum analysis remain.  
We then examine through numerical means the implications of the results for designing
the ideal experiment that would be most sensitive to terms linear in $\theta_{13}$.
Finally, we note that two existing three-neutrino analyses \cite{ggth23,ggpostks} might demonstrate
the correlation between $\theta_{13}$ and $\theta_{23}$.

\section{Vacuum oscillations}
We shall assume CP is conserved. This assumption is indeed significant as all terms in the
oscillation probabilities which are linear in $\sin \theta_{13}$ will be modulated by either $\cos
\delta$ or $\sin \delta$. Thus if CP is violated, our results will be 
modified at the quantitative level. However, in the absence of any evidence for CP
violation in the neutrino sector, we will default to the simplest case of no violation.

In vacuo, the relativistic limit of the neutrino evolution equation in 
the flavor basis is
\begin{equation}
i \partial_t \nu_f = \frac{1}{2E} U \mathcal{M} U^\dagger \nu_f
\end{equation}
for neutrino energy $E$, mixing matrix $U$, and mass-squared matrix $\mathcal{M} = 
\mathrm {diag} (0, \Delta_{21}, \Delta_{31})$.  This first order differential equation is
easily solved to give explicit expressions for the oscillation probabilities. The
probability that a neutrino of flavor $\alpha$ will be detected a distance
$L$ from the source as a neutrino of flavor $\beta$ is
\begin{equation}
\mathcal{P}_{\alpha \beta}(L/E) = \delta_{\alpha \beta} - 4 
\sum^3_{\genfrac{}{}{0pt}{}{k <j}{j,k=1}} (U_{\alpha j} U_{\alpha k} U_{\beta k} 
U_{\beta j}) \sin^2 {\varphi_{jk}}
\end{equation}
with $\varphi_{jk} := \Delta_{jk} L/4E$.  As we are interested in the
oscillatory region for the small mass-squared difference $\Delta_{21}$, we
will assume that the oscillations due to the two larger mass-squared
differences are incoherent; that is, we take
\begin{equation}
\langle \sin^2{\varphi_{31}} \rangle = \langle \sin^2{\varphi_{32}} \rangle
= \frac{1}{2}.
\end{equation}
The $\nu_e$--$\nu_e$ oscillation channel exhibits no linear dependence on
$\sin{\theta_{13}}$, so we neglect it in our analysis.  Using the
shorthand notation $s_{jk}:=\sin{\theta_{jk}}$ and 
$c_{jk}:=\cos{\theta_{jk}}$, the interesting oscillation channels  
in the prescribed limit become
\begin{eqnarray}
\mathcal{P}_{e \mu} &=& [\tfrac{1}{2} \sin{2\theta_{12}} \cos{2\theta_{12}} 
\sin{2\theta_{13}} c_{13} \sin{2\theta_{23}}
\nonumber \\
&& + \sin^2{2 \theta_{12}} c_{13}^2(c_{23}^2-s_{13}^2 s_{23}^2)] \sin^2 
{\varphi_{12}} \nonumber \\ 
&& + \tfrac{1}{2}\sin^2{2\theta_{13}} s_{23}^2,  \label{pem}
\end{eqnarray}
\begin{eqnarray}
\mathcal{P}_{\mu \mu} &=& 1-[\cos^2{2 \theta_{12}} s_{13}^2 \sin^2{2 
\theta_{23}}  \nonumber \\ 
&& + 2 \sin{2 \theta_{12}} \cos{2 \theta_{12}}
s_{13} \sin{2 \theta_{23}} (c_{23}^2 - s_{13}^2 s_{23}^2)\nonumber \\
&& +\sin^2{2\theta_{12}}(c_{23}^2 - s_{13}^2 s_{23}^2)^2] \sin^2 
{\varphi_{12}}   \nonumber \\
&&   -2 c_{13}^2 s_{23}^2 (1-c_{13}^2 s_{23}^2), \label{pmm}
\end{eqnarray}
\begin{eqnarray}
\mathcal{P}_{\mu \tau} &=& [-\sin^2 {2\theta_{12}}(c_{23}^2-s_{13}^2 
s_{23}^2)(s_{23}^2 -s_{13}^2
c_{23}^2) \nonumber \\
&& + \sin{2\theta_{12}} \cos{2\theta_{12}} s_{13}(1+s_{13}^2) 
\sin{2\theta_{23}}
\cos{2\theta_{23}} \nonumber \\
&& +  \cos^2{2\theta_{12}} s_{13}^2 \sin^2{2 \theta_{23}} ] \sin^2 
\varphi_{12} \nonumber \\
&& + \tfrac{1}{2} c_{13}^4 \sin^2 {2 \theta_{23}} . \label{pmt}
\end{eqnarray}

As stated, for the mass-squared difference $\Delta_{31} = 2 \times 10^{-3}$,
the results of CHOOZ \cite{chooz} restrict the magnitude of $\theta_{13}$ by
\begin{equation}
\vert \theta_{13} \vert < 0.22 \qquad (90\%~\mathrm{C. L.}).
\end{equation}
Additionally, an analysis \cite{ggth23} which aims to determine the octant of 
$\theta_{23}$
indicates a range 
\begin{equation}
0.67 < \theta_{23} < 0.91 \qquad (90\%~\mathrm{C. L.}).
\end{equation}
Permitting negative values of $\theta_{13}$, the allowed deviation of
$\theta_{13}$ from zero is on the same order of magnitude as the allowed
deviation of $\theta_{23}$ from maximal mixing. 
We introduce the parameter
$\varepsilon$ to indicate the octant of $\theta_{23}$ via
\begin{equation}
\theta_{23} = \frac{\pi}{4} +\varepsilon
\end{equation}
so  that negative (positive) $\varepsilon$ indicates that $\theta_{23}$ is in the first (second)
octant.

We expand the {\itshape in vacuo} oscillation probabilities in
Eqs.~(\ref{pem}--\ref{pmt}) to first order in $\theta_{13}$ and 
$\varepsilon$. We utilize the relations
\begin{eqnarray}
s_{23}^2 = \frac{1}{2} (1 + \sin 2\varepsilon), && \sin {2 \theta_{23}} =
\cos {2 \varepsilon} ,\\
c_{23}^2 = \frac{1}{2} (1 - \sin 2\varepsilon), && \cos {2 \theta_{23}} =
-\sin {2 \varepsilon}.
\end{eqnarray}
Beginning with the
$\nu_\mu$--$\nu_\tau$ oscillation channel, we find that the 
term linear in $\theta_{13}$ is
effectively screened due to the near maximal
mixing of $\theta_{23}$.  
The result is a modified two-neutrino oscillation
probability
\begin{equation}
\mathcal{P}_{\mu \tau} \simeq \frac{1}{2} -
\frac{1}{4}\sin^2{2\theta_{12}} \sin^2{\varphi_{21}}. \label{pmtlim}
\end{equation}
The Super-K atmospheric experiment indicates that
$\nu_\mu$--$\nu_\tau$ occur most prominently at the $\Delta_{31}$ mass
scale.  With this in mind, we anticipated this result.

From Ref.~\cite{th13exp}, one finds that in the 
oscillation channels $\mathcal{P}_{e
\mu}$ and $\mathcal{P}_{\mu \mu}$ terms linear in $s_{13}$ are
not overly screened.  It is these same terms which are responsible,
in part, for the possible excess of $e$-like events in the sub-GeV
sample of the Super-K atmospheric data \cite{smirnov1}.
Dropping
higher order terms, we find that the $\nu_e$--$\nu_\mu$ channels exhibit
the behavior
\begin{equation}
\mathcal{P}_{e \mu} \simeq \sin^2{2 \theta_{12}} \left[ \frac{1}{2}-
\varepsilon +  \cot({2 \theta_{12}}) \theta_{13} \right] \sin^2
\varphi_{21}. \label{vacpem}
\end{equation}
For $\nu_\mu$--$\nu_\mu$ oscillations, the probability becomes
\begin{equation}
\mathcal{P}_{\mu \mu} \simeq \frac{1}{2}- \sin^2{2\theta_{12}} \left[ 
\frac{1}{4}
 - \varepsilon
+  \cot({2 \theta_{12}}) \theta_{13} \right] \sin^2
\varphi_{21}. \label{vacpmm}
\end{equation}
The effect of $\theta_{13}$ upon these channels is largest whenever the
dynamical term $\sin^2 \varphi_{21}$ is maximal.  At this functions first
peak, one has a value of $L/E = 2 \pi/\Delta_{21} \sim 1.6 \times 10^4$
km/GeV.  This is the result found in Ref.~\cite{th13exp}. There we suggested 
that an ideal experiment would be to avoid the MSW effect by limiting the 
energy to a very low value, say $E< 50$ MeV. This is far from possible with 
existing technology, so we expand the discussion to include matter effects.

\section{Matter oscillations}
We consider neutrinos which traverse a constant density
mantle.  By doing so, we avoid the issue of parametric resonances for
neutrinos which go through the core \cite{parametric}.  A constant density
mantle is a relatively good approximation; however, it limits our baseline
to about 4000 km \cite{barger8}.  This, in turn, places a limit upon the
neutrino energy if we are to reach the peak of the $\Delta_{21}$
oscillations.  In what follows, we consider only neutrino energies
below 1 GeV and assume a constant average mantle density of 3.5
g/cm$^3$ \cite{prem,lisi}. 

For neutrinos traveling through matter, one needs to
modify the Hamiltonian to include an effective potential which operates on 
the electron flavor exclusively \cite{msw}.  In the flavor basis, the 
neutrino evolution equation in matter becomes
\begin{equation}
i \partial_t \nu_f = \left[ \frac{1}{2E} U \mathcal{M} U^\dagger + 
\mathcal{V} \right]\nu_f \label{mswev}
\end{equation}
where the operator $\mathcal{V}$ operates on the electron flavor with a 
magnitude $V= \sqrt{2} G_F N_e$, where $G_F$ is the Fermi coupling constant 
and $N_e$ is the electron number density.  For a mantle density of 3.5 g/cm$^3$, 
this yields a potential around $V \sim 1.3 \times 10^{-13}$ eV.
We note that for anti-neutrinos, we need to change the algebraic sign of this 
potential.  We shall consider only neutrinos below.  

For sub-GeV neutrinos traversing the earth, 
matter affects are most easily addressed
in the propagation basis developed in Ref.~\cite{smirnov1}.  
We will briefly review their derivation.  In the absence of CP violation, the 
vacuum mixing matrix $U$ can be written as the product of three rotations
\begin{equation}
U = U_1(\theta_{23}) U_2(\theta_{13}) U_3(\theta_{12})
\end{equation}
where $U_j(\theta)$ is a proper rotation by angle $\theta$ about the $j$-th 
axis.  
As $U_1(\theta)$ commutes with $\mathcal{V}$, we
may rewrite the evolution equation
\begin{equation}
i \partial_t \nu' =
\left[ \frac{1}{2E} U_3(\theta_{12}) \mathcal{M}
U_3(\theta_{12})^\dagger + U_2(\theta_{13})^\dagger \mathcal{V} U_2 
(\theta_{13}) \right] \nu'
\end{equation}
with $\nu' = U_2(\theta_{13})^\dagger U_1(\theta_{23})^\dagger \nu_f$.

By conjugating the Hamiltonian in this basis via a locally defined 
$U_2(\theta)$, this
new propagation basis can be approximately described
by a Hamiltonian $\widetilde{H}$ which is block diagonal.  This correction to
$\theta_{13}$ is given by
\begin{equation}
\tan {2 \theta} = \frac{2 \sin{2\theta_{13}}E V}{\Delta_{31} -s_{12}^2 
\Delta_{21}
- 2\cos{2\theta_{13}}EV}.
\end{equation}
As $\Delta_{31}$ dominates the denominator and $2EV/\Delta_{31} < 0.13$,
one effectively has
\begin{equation}
\theta \simeq \frac{ \sin{2\theta_{13}}E V}{\Delta_{31}}.
\end{equation}
This correction results in a modified mixing angle
\begin{equation}
\theta_{13}^m = \theta_{13} + \theta.
\end{equation}
With this additional rotation, we define locally the propagation basis with
$\widetilde{\nu} = U_2(\theta)^\dagger \nu'$ and Hamiltonian $\widetilde{H}$
\begin{equation}
\widetilde{H} = \left(
\begin{array}{cc}
H_2 & 0\\
0 & \Delta_{31}/2E + s_{13}^2 V
\end{array}
\right),
\end{equation}
where the block is given by
\begin{equation}
H_2 = \frac{1}{2E} U_3(\theta_{12}) \left(
\begin{array}{cc}
0 & 0\\
0 & \Delta_{21}
\end{array}
\right) U_3(\theta_{12})^\dagger + c_{13}^2 \left(
\begin{array}{cc}
V & 0\\
0 & 0
\end{array}
\right).
\end{equation}

As the correction $\theta$ depends upon the local density, we must consider 
its
temporal (spatial) derivative in the evolution equation.  Using the product 
rule, we note
\begin{equation}
\partial_t \widetilde {\nu} = \partial_t [U_2(\theta)^\dagger] \nu'+
U_2(\theta)^\dagger \partial_t \nu'.
\end{equation}
Letting $\sigma$ be the generator of the rotation so that $U_2(\theta) =
e^{i \theta \sigma}$, we have
\begin{equation}
\partial_t U_2(\theta)^\dagger = -i \sigma U_2(\theta)^\dagger \partial_t
\theta.
\end{equation}
Dropping insignificant terms, one may write the evolution equation in the
propagation basis as
\begin{equation}
i \partial_t \widetilde {\nu} = (\widetilde{H} + \sigma \partial_t \theta) 
\widetilde {\nu}
\end{equation}

Given the neutrino energies and baselines under consideration in our analysis, 
the term $\delta_t \theta$ is negligible.  For our
purposes, it is sufficient to consider the evolution equation
\begin{equation}
i \partial_t \widetilde {\nu} = \widetilde{H} \widetilde {\nu}. \label{prop2}
\end{equation}
The block $H_2$ in this Hamiltonian can be easily diagonalized in closed form
with eigenvalues $\lambda_\pm$.  Of dynamical 
relevance is the difference in these
eigenvalues which yields the effective constant density mass-squared 
difference
\begin{equation}
\Delta_{21}^m = \Delta_{21} \sqrt{\cos^2 {2\theta_{12}}(1-E/E_R)^2 + 
\sin^2{2\theta_{12} }}, \label{d21m}
\end{equation}
where we have defined the resonance energy to be
\begin{equation}
E_R = \frac{\Delta_{21}\cos 2\theta_{12}}{2 V c_{13}^2}.  
\end{equation}
Fixing the solar mixing angle $\theta_{12} = 0.58$, we find the resonance 
energy
in the mantle is $E_R \sim 120$ MeV.
Matter effects require an accommodation to the
other mass-squared difference $\Delta_{31}^m = \Delta_{31}-2E\lambda_-$.
The mixing angle which achieves this diagonalization satisfies
\begin{equation}
\sin 2\theta_{12}^m = \frac{\sin 2 \theta_{12}}{\sqrt{\cos^2 2\theta_{12} (1- 
E/E_R)^2 + \sin^2 2 \theta_{12} }}. \label{th12m}
\end{equation}

With this
diagonalization, we see that, in the flavor basis, Eq.~(\ref{prop2})
becomes
\begin{equation}
i \partial_t \nu_f = \frac{1}{2E} U' \mathcal{M}' U^{'\dagger} \nu_f
\end{equation}
with the modified mass-squared matrix $\mathcal{M}'=(0, \Delta_{21}^m, 
\Delta_{31}^m)$.  The new mixing matrix accounts for the matter mixing angles
\begin{equation}
U' = U_1(\theta_{23}) U_2(\theta_{13}^m) U_3(\theta_{12}^m).
\end{equation}

\begin{figure}
\includegraphics[width=3in]{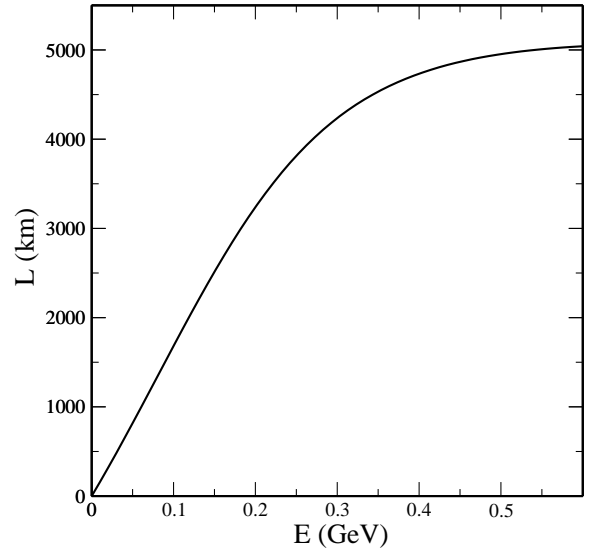}
\caption{Ideal baseline for mono-energetic neutrinos traveling through
a constant density mantle.
\label{baseline}}
\end{figure}

We now discuss these modifications within the context of incoherent (coherent)
oscillations of the large (small) mass-squared difference.  As the matter
potential $V$ is larger than the kinetic term $\Delta_{21}/2E$ for the 
energies of pertinence, the effect on the smaller mass-squared difference is
most profound.  Examining Eq.~(\ref{d21m}), we note that for neutrinos of
energies below twice the resonance energy, the effective mass-squared
difference, and thus frequency of oscillations, decreases.  For neutrino
energies greater than twice this resonance energy, the frequency increases. 
Assuming a mono-energetic neutrino source, the first peak of $\sin^2
\varphi^m_{21}$ occurs at a baseline
\begin{equation}
L = 2\pi E/ \Delta_{21}^m.
\end{equation}
Setting $\theta_{12}=0.58$, we see in Fig.~(\ref{baseline}) how this ideal
baseline changes as a function of energy.
Recalling that the resonance energy is 120 MeV, we find that at this energy a baseline
around 2000 km will yield maximal sensitivity to the terms linear in
$\theta_{13}^m$.  At twice the resonant energy, the necessary baseline has
increased to 3700 km.  Finally, the maximal baseline permitted by our constant
density approximation is 4000 km, which corresponds to a neutrino energy of 270 MeV.

Regarding the larger mass-squared difference $\Delta_{31}$, the modification of
this quantity needed to account for matter was the subtraction of $2 E \lambda_-$.  This
term is at most on the order of $10^{-4}$ eV$^2$; therefore, our 
assumptions regarding the incoherence of these oscillations is unaffected.  
In short, we may still take
\begin{equation}
\langle \sin^2 \varphi_{31}^m \rangle = \frac{1}{2}.
\end{equation}

With respect to the mixing angles, the mixing angle $\theta_{12}$ undergoes 
the
greatest degree of modification to accommodate matter effects.  This 
discussion 
is given in the next section.  The effect upon
$\theta_{13}$ is quite small.  Estimating the maximum value of the additive
correction, we have
\begin{equation}
\theta \simeq \frac{2 E V}{\Delta_{31}} \theta_{13} \lesssim 
0.09~\theta_{13}.
\end{equation}
This represents at most ten percent correction for neutrinos of energies around
1 GeV.  Should we only consider neutrino energies below 300 MeV, the correction
becomes less than three percent.  We shall consider this correction in our
numerical analysis but drop the term in the analytical discussion. 
Incidentally, this correction to $\theta_{13}$ is the only place where the sign
of $\Delta_{31}$ has any effect.  Hence, we see that the hierarchy ambiguity
plays a very small role in these calculations.  Finally, the remaining mixing
angle $\theta_{23}$ is insensitive to matter effects.

The upshot is that Eqs.~(\ref{pem})--(\ref{pmt}) are still valid for 
neutrinos
traveling through a constant density medium provided we make the appropriate
substitutions for the mixing angles and small mass-squared difference.
Neglecting the small correction $\theta$, the first order expansion of these
probabilities about $\theta_{13}$ and $\varepsilon$ have not changed in form
\begin{equation}
\mathcal{P}_{e \mu}^m \simeq  \sin^2{2\theta_{12}^m} \left[ \frac{1}{2} 
-\varepsilon
+ \cot({2 \theta_{12}^m}) \theta_{13} \right] \sin^2 \varphi_{21}^m\,, 
\label{pem1stm}
\end{equation}
and for $\nu_\mu$--$\nu_\mu$ oscillations, the probability becomes
\begin{equation}
\mathcal{P}_{\mu \mu}^m \simeq \frac{1}{2}- \sin^2{2\theta_{12}^m}  \left[ 
\frac{1}{4}
- \varepsilon 
+ \cot({2 \theta_{12}^m}) \theta_{13} \right] \sin^2 \varphi_{21}^m. 
\label{pmm1stm}
\end{equation}

\section{Discussion}

\begin{figure}
\includegraphics[width=3in]{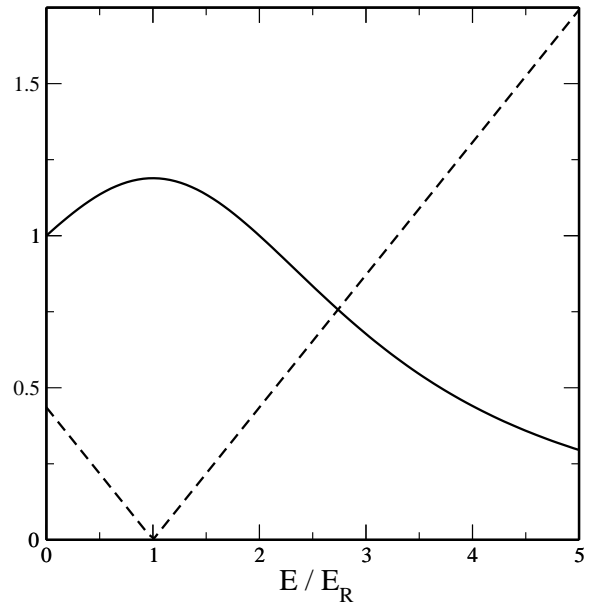}
\caption{The solid line is the ratio $\sin^2 2\theta_{12}^m/\sin^2
2\theta_{12}$.  The dotted line is the curve $\cot 2 \theta_{12}^m$.
We have set $\theta_{12}=0.58$ and plot the curves as functions
of the dimensionless ratio $E/E_R$.
\label{th12mfig}}
\end{figure}

Aside from the adjustment in the ideal baseline, the change in the
$\theta_{12}^m$ mixing angle is the most critical 
modification needed to account for matter effects.  
In the approximate formulae in Eqs.~(\ref{pem1stm}) and (\ref{pmm1stm}), the 
oscillatory
terms are modulated by $\sin^2 2\theta_{12}^m$ thereby affecting the 
amplitude
of oscillations.
Recalling
Eq.~(\ref{th12m}), we note that, relative to vacuum oscillations, this factor 
enhances the oscillations in matter provided that the neutrino energy is less
than twice the resonant energy.
The new matter mixing angle is maximal whenever $E \simeq E_R$ and equal to 
the vacuum value whenever $E \ll E_R$ or $E \simeq 2 E_R$.  For neutrino 
energies greater than twice the resonant energy, the factor $\sin^2 
2\theta_{12}^m$ decreases significantly, effectively suppressing the 
amplitude of oscillations.  In Fig.~(\ref{th12mfig}), 
we plot the ratio $\sin^2 2 \theta_{12}^m/\sin^2 2 \theta_{12}$ as a function 
of
neutrino energy in order to demonstrate the enhancement and suppression of the
oscillation amplitude relative to vacuum oscillations. At an energy of 270 
MeV,
which is $2.25 E_R$, we find that the amplitude will be suppressed by a 
factor
of 0.9  relative to vacuum oscillations.

Another notable feature of the first order expansions of the oscillation
channels in Eqs.~(\ref{pem1stm}) and (\ref{pmm1stm}) are the terms which 
appear 
within the square brackets.  Given the lack on knowledge in the true values 
of
$\theta_{23}$ and $\theta_{13}$, it is apparent from the equations that a
measurement at these baselines of, say, the $\nu_e$--$\nu_\mu$ probability does 
not
directly resolve the values of these mixing angles.  Indeed, we see that a 
host
of parameters can satisfy a measurement provided the linear relation 
\begin{equation}
\varepsilon
-  \cot({2 \theta_{12}^m}) \theta_{13} = \mathrm{const}
\end{equation}
holds.
  
As the slope of this relation, $\cot {2 \theta_{12}^m}$, varies with neutrino
energy, its value is represented by the dotted curve in 
Fig.~(\ref{th12mfig}).
For energies greater than the $E_R$, the slope increases linearly with 
neutrino
energy.  Of particular importance are two points.  First, at the resonance
energy we see that both oscillation channels do not depend linearly on
$\theta_{13}$; a measurement of such neutrinos could at best improve the 
bounds
upon $\theta_{23}$ while potentially resolving the octant of $\theta_{23}$. 
As our constant density approximation is really only valid up to 270 MeV, we
note from the graph the value of $\cot 2\theta_{12}^m$ is 0.54 at this energy.
A measurement at this energy would have implications for both the sign of
$\theta_{13}$ and the octant of $\theta_{23}$.

The aim of this paper is to determine how a measurement at this baseline
of either the $\nu_e$--$\nu_\mu$ or $\nu_\mu$--$\nu_\mu$ oscillation channel will affect the
values of $\theta_{23}$ and $\theta_{13}$.  To most explicitly demonstrate 
this,
we will create some mock experimental data and combine it with the present
knowledge of these mixing angles.  To this end, we introduce a $\chi^2$ 
function
made up of three parts
\begin{equation}
\chi^2 = \chi^2_{13} + \chi^2_{23} + \chi^2_\mathcal{P}.
\end{equation}
For the first term, we include information from CHOOZ via the simple relation
\begin{equation}
\chi^2_{13} = \frac{(\theta_{13}-\theta_{13}^0)^2}{\sigma_{13}^2}
\end{equation}
where the best fit value is $\theta_{13}^0=0$ and the standard deviation is
$\sigma_{13} = 0.15$.   Similarly, we include information on $\theta_{23}$ in 
the same manner where the best fit value is maximal $\theta_{23}^0 = 0.785$ 
and the
standard deviation is $\sigma_{23} = 0.07$.

The final term in $\chi^2$ represents our mock data.  For concreteness, we 
choose
the electron-muon oscillation channel.  As with the others, we define
$\chi^2_\mathcal{P}$ with respect to a best fit $\mathcal{P}_{e\mu}^0$ and a
standard deviation which we take to be ten percent of the best fit
\begin{equation}
\chi^2_\mathcal{P} = \frac{(\mathcal{P}_{e\mu}(\theta_{13},\theta_{23})
-\mathcal{P}_{e\mu}^0)^2}{\sigma_{\mathcal{P}}^2}
\end{equation} 
We assume, as well, a mono-energetic source of 270 MeV measured at the ideal
baseline of 4000 km.
The question remains as to what value we assign our measured best fit.
At present, the only experimentally observed neutrinos of this energy and 
baseline occur in the sub-GeV sample of the Super-K atmospheric experiment 
\cite{superk}.  A detailed analysis of the oscillation channels of interest 
has been performed in Ref.~\cite{smirnov1}.  Our first order approximations 
in $\varepsilon$ and $\theta_{13}$ exhibit some of the significant behavior 
given in this reference.  
Briefly, assuming neutrino oscillations and CP invariance, one would expect 
the detector-to-source ratio of atmospheric 
electron-like neutrinos to be
\begin{equation}
\mathcal{R}_e = \mathcal{P}_{ee} + r \mathcal{P}_{e \mu},
\end{equation}
where $r$ is the ratio of muon to electron neutrinos in the upper atmosphere.
For energies of a few hundred MeV, one has $r \simeq 2$.  
For those
which traverse the mantle only with baselines over $3 \times 10^3$ km, we may 
use the expression from Eq.~(\ref{pem1stm}) along with a first order 
expansion of $\mathcal{P}_{ee}$ to determine the approximate expectation ratio
\begin{equation}
\mathcal{R}_e \simeq 1+ 2 \sin^2 2 \theta_{12}^m \left[ 
 \cot({2 \theta_{12}^m}) \theta_{13} - \varepsilon 
 \right] \sin^2 \varphi_{21}^m. \label{re}
\end{equation}

In-depth analyses of Super-K \cite{smirnov1,eexcess,ggth23} 
cite the oscillations discussed herein as
a potential explanation for an excess of electron-like events in the sub-GeV
sample.  If this excess is indeed real, then our approximation indicates
\begin{equation}
\cot({2 \theta_{12}^m}) \theta_{13} - \varepsilon >0. \label{excess}
\end{equation}
Though this expression is not the result of a detailed analysis of the
experiment, its qualitative 
features should still be valid.  
With that said, the inequality in (\ref{excess}) 
indicates that the potential excess of electrons in the sub-GeV sample can 
be accounted for by a range of parameters.  Should $\theta_{13}=0$ for 
instance, we find that $\varepsilon <0$ which indicates that $\theta_{23}$ is 
in the first octant.  Alternatively, should 2--3 mixing be maximal, the 
above 
inequality indicates that $\theta_{13}$ is positive.

This qualitative understanding is confirmed in more detailed analyses of experiments which have 
very long baselines.
In Ref.~\cite{ggth23}, preliminary evidence indicates that $\theta_{23}$ lies 
in the 
first octant.  Within this full three-neutrino analysis (labeled A), the 
best fit mixing angles are found to be
\begin{equation}
\theta_{12}^A=0.575, \qquad \theta_{13}^A=0.0, \qquad \theta_{23}^A= 0.745,
\label{seta}
\end{equation}
which in the present context is expressed by $\varepsilon = -0.04$.
Additionally, in Ref.~\cite{ggpostks}, a full three-neutrino analysis of 
the world's data is performed.
When 2--3 mixing is assumed maximal, 
the best fit mixing angles (labeled B) are determined to be
\begin{equation}
\theta_{12}^B=0.591, \qquad \theta_{13}^B=0.095, \qquad \theta_{23}^B= 0.785.
\end{equation}
These two parameter sets are consistent with the correlation between $\theta_{13}$ and 
$\varepsilon$ which we have indicated above.  Employing the average 
of the two values for the solar mixing angle $\overline{\theta}_{12}=0.58$, we
find 
\begin{equation}
\frac{\theta_{23}^B - \theta_{23}^A}{\theta_{13}^B - \theta_{13}^A} = 
\cot 2\overline{\theta}_{12},
\end{equation}
to four percent. Thus, although a small effect, the contribution to the $e$-like sub-GeV
excess for atmospheric data \cite{smirnov1} introduces a correlation between the extracted 
values of $\theta_{13}$ and $\varepsilon$.

In Ref.~\cite{ggth23} it is shown that future atmospheric data will further constrain the
value of $\theta_{23}$ and perhaps determine its octant, this for $\theta_{13}=0$. We 
have here implicitly generalized this result. Future atmospheric data will provide information 
on the value of the function $\cot(2\theta_{12}^m)\,\theta_{13}-\varepsilon$ in addition to the
cleaner measurements of $\varepsilon$.

\begin{figure}
\includegraphics[width=3in]{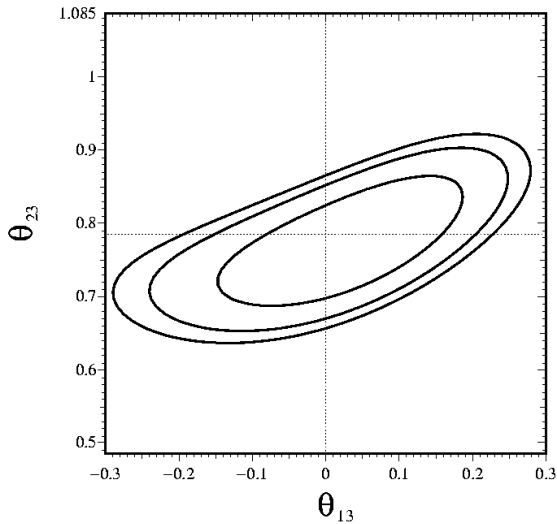}
\caption{Given the mock data for the $\mathcal{P}_{e
\mu}$ measurement at the ideal baseline for mono-energetic neutrinos of 270 
MeV,
we construct a hypothetical allowed region for $\theta_{13}$ and 
$\theta_{23}$. 
We draw the 68\%, 90\%, and 95\% CL contours.
\label{cl}}
\end{figure}

Though these values are not statistically robust, we may use them as
an indication of where a best fit point might lie.  For the best fit, we choose
the average value for the solar mixing angle and the values for $\theta_{13}$ and
$\theta_{23}$ from parameter set A in Eq.~(\ref{seta}).  We may now determine 
the acceptable regions for these two mixing angles given our mock data and the
existing knowledge.  From the $\chi^2$ function, we are able to plot the
68\%,
90\%, and 95\% CL allowed regions in Fig.~(\ref{cl}). 
The best fit value is $\theta_{13}=0.04$ and $\theta_{23}=0.775$.  The slope 
of 
the upper portion of the
closed contours in the figure is given by $\cot{2\theta_{12}^m}$.  We note 
that this
mock data indicates a preference for positive $\theta_{13}$ and first-octant 
$\theta_{23}$; however, the possibility remains that they lie elsewhere.  

\section{Conclusion}

We have expanded the oscillation probabilities $\mathcal{P}_{e \mu}$ and 
$\mathcal{P}_{\mu \mu}$ to first order in $\theta_{13}$ and $\varepsilon$, 
where $\varepsilon$ indicates the deviation of the mixing angle $\theta_{23}$ from 
maximal. This was done for neutrinos propagating in the vacuum and in a constant 
density mantle.  For the oscillations in matter, it was shown that the 
amplitude of the oscillations is enhanced relative to vacuum for neutrino
energies less than twice the resonant energy; the amplitude was suppressed at 
energies greater than this mark.  Terms linear in $\theta_{13}$ were of 
significance whenever the matter-modified solar mass-squared difference 
$\Delta_{21}^m$ was in the oscillatory region.  The ideal baseline, defined 
by the first peak or trough of these oscillations, was determined as a 
function of neutrino energy.  For a maximum baseline of 4000 km, this energy 
was found to be 270 MeV.  At this energy and ideal baseline, the relation 
between the deviation of $\theta_{23}$ from maximal mixing and $\theta_{13}$ 
was examined by creating mock data.  The mock data demonstrated what was 
found in the approximate equations; that is, $\theta_{13}$ and 
$\varepsilon$ are linearly related via a slope of $\cot{2 \theta_{12}^m}$.  
The energy dependence of this slope was elucidated.  For an energy of 270 
MeV, it was shown to be 0.54; this factor increases linearly with neutrino 
energy, assuming the energy is greater than the resonant energy.  

It is seen from the mock data that this ideal measurement could help 
determine the algebraic sign of $\theta_{13}$ and perhaps the octant of 
$\theta_{23}$.  For a measurement most sensitive to the octant of 
$\theta_{23}$, one would like the neutrino energy to be near the MSW 
resonance of 120 MeV.  At this point, the slope is minimal because $\cot{2 \theta_{12}^m}=0$.   
To best determine the sign of $\theta_{13}$, one would like the slope 
relating the mixing angle and $\varepsilon$ to be as large as possible 
without completely suppressing the oscillation amplitude.  To achieve this, 
one would require higher energy neutrinos and, therefore, longer baselines.  
Our constant density treatment would no longer be valid as these neutrinos 
would certainly traverse the core of the earth and parametric resonances 
would be of consequence.  We leave a detailed analysis of this case to future work.

\acknowledgements\acknowledgements

Work supported, in part, by the US Department of Energy under contract DE-FG02-96ER40975. 
The authors thank S.\ Palomares-Ruiz for helpful conversations.

\bibliography{th13th23b}

\begin{thebibliography}{1}

\bibitem{lsnd}A. Aguilar {\it et al.}, Phys. Rev. D {\bf 64}, 112007 (2001).

\bibitem{pdg}
Particle Data Group, Phys. Lett. {\bf B592} (2004).

\bibitem{solar}
B.~T.~Cleveland {\it et al.}, Astrophys. J. {\bf 496}, 505 (1998);
J.~N.~Abdurashitov {\it et al.}, Phys. Rev. C {\bf 60}, 055801 (1999);
J. Exp. Theor. Phys. {\bf 95}, 181 (2002);
W.~Hampel {\it et al.}, Phys. Lett. {\bf B447}, 127 (1999);
M.~Altmann {\it et al.}, Phys. Lett. {\bf B490}, 16 (2000);
Q.~R.~Ahmad {\it et al.}, Phys. Rev. Lett. {\bf 87}, 071301 (2001);
Phys. Rev. Lett. {\bf 89}, 011301 (2002);
S.~N.~Ahmed, Phys. Rev. Lett. {\bf 92}, 181301 (2004).

\bibitem{kamland}
K.~Eguchi {\it et al.} [KamLAND Collaboration], Phys. Rev. Lett. {\bf 90}, 
021802 (2003); 

T.~Araki  {\it et al.} [KamLAND Collaboration], Phys. Rev. Lett. {\bf 94}, 
081801 
(2005).

\bibitem{superk}
Y. Fukuda {\it et al.} [Super-Kamiokande Collaboration],
Phys. Lett. {\bf B335}, 237 (1994); {\bf B433}, 9 
(1998);
{\bf B436}, 33 (1998); Phys. Rev. Lett. {\bf 81}, 1562 (1998);
Phys. Lett. {\bf B436}, 33 (1998); Phys. Rev. Lett. {\bf 82}, 2644 
(1999); {\bf 86}, 5651 (2001); Y. Ashie {\it et al.} [Super Kamiokande Collaboration],
Phys. Rev. Lett. {\bf 93}, 101801 (2004); hep-ex/0501064.

\bibitem{k2k}
M.~H.~Ahn {\it et al.} [K2K Collaboration], 
Phys. Rev. Lett. {\bf 90}, 041801 (2003);
Phys.\ Rev.\ Lett.\  {\bf 93}, 051801 (2004);
E.~Aliu {\it et al.}  [K2K Collaboration], 
Phys.\ Rev.\ Lett.\  {\bf 94}, 081802 (2005).

\bibitem{chooz}
M. Apollonio {\it et al.}, Phys. Lett. {\bf B 420}, 397 (1998); {\bf B466}, 
415 (1999); Eur. Phys. J. {\bf C 27}, 331 (2003).

\bibitem{MSTV} 
M. Maltoni, T. Schwetz, M. T\'ortola, and J. W. F. Valle, New J. Phys.
{\bf 6}, 122 (2004).

\bibitem{BGP} 
J. N. Bahcall, M. C. Gonzalez-Garcia, and C. Pe\~na-Garay, JHEP {\bf 0408}, 
016 (2004)

\bibitem{bargerrev}
V. Barger, D. Marfatia, and K. Whisnant, Int. J. Mod. Phys. E {\bf 12}, 
569 (2003).

\bibitem{neutreview}
  M.~C.~Gonzalez-Garcia and Y.~Nir, Rev.\ Mod.\ Phys.\  {\bf 75}, 345 (2003).

\bibitem{BP} 
J. N. Bahcall and C. Pe\~na-Garay, New J. Phys. {\bf 6}, 63 (2004).

\bibitem{BK} B. Kayser, hep-ph/0506165; G. L. Fogli, E. Lisi, A. Marrone, and A. Palazzo,
hep-ph/0506083
  
\bibitem{reactor}
K.~Anderson {\it et al.}, hep-ex/0402041.

\bibitem{sergio1} 
J. Bernb\'eu and S. Palomares-Ruiz, JHEP {\bf 0402}, 068 (2004).

\bibitem{sergio2} 
J. Bernb\'eu, S. Palomares-Ruiz, and S. T. Petcov, Nucl. Phys.
{\bf B669}, 255 (2003).

\bibitem{sergio3} 
S. Palomares-Ruiz and S. T. Petcov, Nucl. Phys. {\bf B712}, 392 (2005).

\bibitem{DIMVNM} 
D. Indumathi and M. V. N. Murthy, Phys. Rev. D {\bf 71}, 013001 (2005).

\bibitem{minos}
R.~Saakian  [MINOS Collaboration], Phys. Atom. Nucl.  {\bf 67}, 1084 (2004)
[Yad. Fiz.  {\bf 67}, 1112 (2004)].

\bibitem{angles}
D.~C.~Latimer and D.~J.~Ernst, Phys.\ Rev.\ D {\bf 71}, 017301 (2005).

\bibitem{expansion}
E. K. Akhmedov, R. Johansson, M. Lindner, T. Ohlsson, and T. Schwetz,
JHEP {\bf 0404}, 078 (2004).

\bibitem{smirnov1}
O. L. G. Peres and A. Yu. Smirnov, Nucl. Phys. B {\bf 680}, 479 (2004).

\bibitem{eexcess}
C. W. Kim and U. W. Lee, Phys. Lett. {\bf B444}, 204 (1998);
O. L. G. Peres and A. Yu. Smirnov, Phys. Lett. {\bf B456}, 204
(1999).

\bibitem{CGMM} M. C. Gonzalez-Garcia and M. Maltoni, Eur. Phys. J. C {\bf 26}, 417 (2003).

\bibitem{th13exp}
D. C. Latimer and D. J. Ernst, Phys. Rev. C {\bf 71}, 062501(R) (2005).

\bibitem{us} D. C. Latimer and D. J. Ernst, nucl-th/0404059.

\bibitem{msw}
L.~Wolfenstein, Phys.\ Rev.\ D {\bf 17}, 2369 (1978);
S.~P.~Mikheev and A.~Y.~Smirnov, Sov.\ J.\ Nucl.\ Phys.\  {\bf 42}, 913 (1985)
[Yad.\ Fiz.\  {\bf 42}, 1441 (1985)].

\bibitem{freund}
M. Freund, P. Huber, and M. Lindner, Nucl. Phys. B {\bf 615}, 331 (2001).

\bibitem{ggth23}
M. C. Gonzalez-Garcia, M. Maltoni, and A. Yu. Smirnov, Phys. Rev. D {\bf 70}, 
093005 (2004).

\bibitem{ggpostks}
M.~C.~Gonzalez-Garcia and C.~Pe\~na-Garay, Phys. Rev. D {\bf 68}, 093003 
(2003).

\bibitem{parametric}
E.~K.~Akhmedov, A.~Dighe, P.~Lipari and A.~Y.~Smirnov,
Nucl.\ Phys.\ {\bf B542}, 3 (1999).

\bibitem{barger8}
V. Barger, D. Marfatia, and K. Whisnant, Phys. Rev. D {\bf 65},
073023 (2002).

\bibitem{prem}
A.~M.~Dziewonski and D.~L.~Anderson, Phys. Earth Planet. Inter.
{\bf 25}, 297 (1981).

\bibitem{lisi}
E. Lisi and D. Montanino, Phys. Rev. D {\bf 56}, 1792 (1997).


\end{thebibliography}

\end{document}